\documentclass[11pt]{article}
\usepackage[left=2cm,right=2cm,vmargin=2cm]{geometry}

\usepackage{amsmath,booktabs,setspace,xcolor,rotating,hyperref,authblk}
\title{Bayesian correction for covariate measurement error: a frequentist evaluation and comparison with regression calibration}

\newcommand{\etal}{\textit{et al }}

\author{Jonathan W. Bartlett}
\affil{Statistical Innovation Group, AstraZeneca \\ Riverside 2, Granta Park, Cambridge, CB21 6GP UK}
\author{Ruth H. Keogh}
\affil{Department of Medical Statistics \\ London School of Hygiene \& Tropical Medicine}

\date{}

\begin{document}

\maketitle

\begin{abstract}
Bayesian approaches for handling covariate measurement error are well established, and yet arguably are still relatively little used by researchers. For some this is likely due to unfamiliarity or disagreement with the Bayesian inferential paradigm. For others a contributory factor is the inability of standard statistical packages to perform such Bayesian analyses. In this paper we first give an overview of the Bayesian approach to handling covariate measurement error, and contrast it with regression calibration (RC), arguably the most commonly adopted approach. We then argue why the Bayesian approach has a number of statistical advantages compared to RC, and demonstrate that implementing the Bayesian approach is usually quite feasible for the analyst. Next we describe the closely related maximum likelihood and multiple imputation approaches, and explain why we believe the Bayesian approach to generally be preferable. We then empirically compare the frequentist properties of RC and the Bayesian approach through simulation studies. The flexibility of the Bayesian approach to handle both measurement error and missing data is then illustrated through an analysis of data from the Third National Health and Nutrition Examination Survey.
\end{abstract}


\section{Introduction}
Many epidemiological studies are affected by measurement error in one or more of the covariates of interest. It is well known that error in covariates results in biased estimates of true covariate(s)-outcome associations and in a loss of power to detect such associations \cite{Carroll2006}. In this paper we focus on correcting for the effects of measurement error in continuous covariates in three models which are commonly used in epidemiological analysis; linear regression models for continuous outcomes, logistic regression models for binary outcomes, and Cox proportional hazards models for survival or time to event outcomes.

Throughout most of the paper we will focus on the situation in which there is one main exposure of interest, which is subject to measurement error, and one or more other covariates to be adjusted for, which are assumed to be measured without error. The variable which is measured with error could equally be one of the confounders, and indeed the approaches we describe also extend to the more general case of multiple covariates measured with error. While exposure measurement error is commonly prioritized, measurement error in confounders is also a serious and highly prevalent issue, and causes estimates of exposure effects to only be partially adjusted for the poorly measured confounder. Error in continuous variables can take a number of forms. The most simple, and most commonly assumed form is the classical measurement error model, under which the measured exposure is equal to the true exposure plus an independent random error term. Under this model, the measured exposure is an unbiased measure of the true exposure. The error terms are assumed to have zero mean and, typically, constant variance.

To make corrections for the effects of covariate measurement error in regression models requires some information about the relationship between the true exposure and the measured exposure, i.e. regarding the parameters of a measurement error model. One way of gaining information about the error model is to use a validation study within the main study sample, in which the true exposure is observed alongside the measured exposure. It is often not feasible or even possible however to obtain a validation sample and a more common alternative is to obtain one or more replicate observations of the measured exposure for a subset of individuals within the main study sample. We refer to this as a replication study. In this paper we focus on replication studies.

Many methods have been described for correcting for the effects of measurement error in regression models \cite{Carroll2006}. The most widely used correction method is regression calibration (RC), which is popular due to its simplicity and applicability in different types of regression models. In RC, the true exposure, which is unobserved in the main study sample, is replaced when fitting the outcome regression model by the expected value of the true exposure, conditional on the measured exposure and the other error-free covariates for each individual. Regression calibration gives consistent estimates of the true associations between the explanatory variables and the outcome in a linear regression model, and approximately consistent estimates in non-linear models, including logistic regression models \cite{Armstrong1985,Rosner1990} and Cox proportional hazards models \cite{Prentice1982}.

Regression calibration has some drawbacks, however. First, for non-linear models estimates can have moderately large biases even when the sample size is large, particularly if the effect size (odds ratio or hazard ratio) is large \cite{Wang1997}. Second, RC does not automatically accommodate uncertainty in the parameters indexing the measurement process. Measures of uncertainty require use of approximate methods (the `delta method' approach), bootstrapping methods, which are computationally intensive, or estimating equation methods, whose validity relies on asymptotic conditions and are complex to implement in practice. Third, extending the basic RC approach to more complex situations, such as when the outcome model is assumed to depend on non-linear functions of the true covariate \cite{Strawbridge2011}, or when the measurement error models is more complex (e.g. heteroscedastic error \cite{Guo2011}) is not trivial.

The Bayesian approach has been often advocated as a natural route to accommodating sources of uncertainty, including measurement error, misclassification, and missing data. Early papers include those by Richardson and Gilks, who described a Bayesian approach to handling measurement error \cite{Richardson1993a,Richardson1993b}. By taking a Bayesian approach to handle covariate measurement error, uncertainty in the parameters indexing the measurement process is automatically accommodated. Like the method of maximum likelihood (ML), the posterior distributions involved typically involve intractable likelihoods, but this difficulty is obviated by Markov Chain Monte Carlo methods, which are now implemented in a number of standard and Bayesian specific packages. A further strength is that these software packages allow one to define and fit quite complex user defined Bayesian models, meaning that there is great flexibility in adapting the modelling assumptions to the situation at hand. Lastly, and in contrast to methods such as ML or multiple imputation, whose inferences typically rely on various large sample assumptions (e.g. to handle nuisance parameters or in deriving simple imputation combination rules), Bayesian methods do not. In the setting of covariate measurement error, estimators which allow for the error typically have skewed sampling distributions, and this is automatically accommodated in a Bayesian approach, since the entire posterior distribution is simulated.

Despite excellent book length treatments of covariate measurement error methods \cite{Carroll2006}, including one specifically focusing on Bayesian methods \cite{Gustafson2003}, in our view it nevertheless continues to be underused by the epidemiological and clinical research communities. This may be for a number of reasons, but principal among them may be the apparent need to move from a frequentist to a Bayesian inferential approach and the fact that standard statistical packages have (with exceptions) not enabled such Bayesian models to be fitted. To the first of these reasons, as has been noted by others (e.g. \cite{Little2011a}), Bayes procedures often have good frequentist properties, and indeed in small samples can have better frequentist properties than ML methods. As such, one may be able to use a Bayesian method without necessarily adopting the Bayesian inferential paradigm. To this end, we present simulation results to examine the frequentist properties of the Bayesian approach to covariate measurement error, using certain default priors. To the second reason, major steps forward have been made over the last 25 years in terms of accessible MCMC software, such that software and computational power are usually not a hinderance to using a Bayesian approach. Moreover, we make all of our code available online to faciliate increased use of the Bayesian approach.

In Section \ref{setupnotation} we begin by describing the assumed setup and notation for the covariate measurement error problem. Next, in Section \ref{rc} we review the regression calibration approach. In Section \ref{bayes} we describe the Bayesian approach, both in terms of modelling choices and statistical properties, and its practical implementation. We contrast the Bayesian approach with ML and multiple imputation in Section \ref{mlmi}. In Section \ref{sims} we evaluate the frequentist properties of RC and Bayesian analysis in a series of simulation studies of the most common outcome model types. In Section \ref{example} we present results of illustrative analyses using data from the Third National Health and Nutrition Examination Survey (NHANES III). We conclude in Section \ref{discussion} with a discussion.

\section{Setup and notation}
\label{setupnotation}
In this section we describe the general setup used for the remainder of the paper.

\subsection{Outcome model}
We assume data are available for an i.i.d. sample of $n$ individuals. For individual $i$, we let $Y_{i}$, $X_{i}$ and $\mathbf Z_{i}$ respectively denote the outcome, true covariate which is subject to measurement error and error-free covariates. We consider three types of outcome models for $Y_{i}$ (i) linear, (ii) logistic, and (iii) Cox proportional hazards regression. We assume that the outcome model includes only main effects of $X_{i}$ and $\mathbf Z_{i}$. For a linear regression outcome model, we thus assume that
\begin{eqnarray*}
Y_{i} = \beta_{0} + \beta_{X} X_{i} + \beta^{T}_{Z} \mathbf Z_{i} + \epsilon_{i}
\label{eq-linear-outcome}
\end{eqnarray*}
where $\epsilon_{i} \sim N(0, \sigma^{2})$ is an independent normally distributed residual error. For a logistic regression outcome model we assume that
\begin{eqnarray}
\mbox{logit}\left\{P(Y_{i}=1|X_{i},\mathbf Z_{i})\right\} = \beta_{0} + \beta_{X} X_{i} + \beta^{T}_{Z} \mathbf Z_{i}
\label{eq-logistic-outcome}
\end{eqnarray}
Lastly, in the case of a censored time to event outcome, the outcome $Y_{i}=(T_{i},D_{i})$ where $T_{i}$ denotes the observed event or censoring time and $D_{i}$ denotes the event indicator. We then assume a Cox proportional hazards outcome model, such that the hazard given $X_{i}$ and $\mathbf Z_{i}$ is given by
\begin{eqnarray*}
h(t|X_{i},\mathbf Z_{i}) = h_{0}(t) \exp(\beta_{X} X_{i} + \beta^{T}_{Z} \mathbf Z_{i})
\end{eqnarray*}
where $h_{0}(t)$ denotes the baseline hazard function. In the standard frequentist analysis based on the Cox proportional hazards model the baseline hazard is left unspecified and inferences about the hazard ratio parameters $(\beta_{X},\beta_{Z})$ are made via a partial likelihood \cite{Cox1972}.

\subsection{Measurement error model}
We assume that for each study individual, an error-prone measurement $W_{i1}$ is available, rather than the covariate of interest $X_{i}$. We assume a classical error model:
\begin{eqnarray*}
W_{i1}=X_{i}+U_{i1}
\end{eqnarray*}
where $E(U_{i1}|X_{i})=0$. We also assume that the errors $U_{i1}$ are independent of all other random variables. This implies that the error is non-differential with respect to the outcome $Y_{i}$.

In order to allow for the error in $W_{i1}$, we assume the existence of an internal replication sub-study. This means that for a randomly selected group of individuals a second error-prone measurement $W_{i2}=X_{i}+U_{i2}$ is obtained, where the error $U_{i2}$ is assumed independent of $U_{i1}$. We let $\mathbf W_{i}$ denote the vector of error-prone measurements on individual $i$, and let $N_{i}$ denote its length, which is two for those individuals in the replication sub-study and one for those not. In the following we specify further assumptions as required by RC and Bayesian methods.

\section{Regression calibration}
\label{rc}
In the simplest version of regression calibration (RC), the outcome model is fitted as usual, with the unobserved $X_{i}$ replaced by an estimate of $E(X_{i}|W_{i1},\mathbf Z_{i})$ \cite{Keogh2014}. Typically the latter conditional expectation is assumed to be linear in $W_{i1}$ and $\mathbf Z_{i}$, and it can be estimated by linearly regressing $W_{i2}$ on $W_{i1}$ and $\mathbf Z_{i}$ in the individuals from the internal replication substudy. We note that this version of RC does \emph{not} rely on an assumption that the two errors $U_{i1}$ and $U_{i2}$ have the same variance.

If one is willing to make additional assumptions, a somewhat more efficient version of RC can be used, in which $X_{i}$ is replaced by an estimate of $E(X_{i}|\mathbf W_{i},\mathbf Z_{i})$, and the parameters involved in the latter are estimated using all study individuals. A common assumption is to assume that $X_{i}|\mathbf Z_{i} \sim N(\gamma_{0} + \gamma^{T}_{Z} \mathbf Z_{i},\sigma^{2}_{X|Z})$ and that the measurement errors $U_{i1}$ and $U_{i2}$ are normally distributed with mean zero and common variance $\sigma^{2}_{U}$. The parameters can be estimated by ML as a random-intercepts mixed model for the $\mathbf W_{i}$ conditional on $\mathbf Z_{i}$. It then follows from standard properties of the multivariate normal distribution that $X_{i}|\mathbf W_{i},\mathbf Z_{i}$ is normally distributed, with
\begin{eqnarray}
E(X_{i}|\mathbf W_{i},\mathbf Z_{i}) &=& \gamma_{0} +\gamma^{T}_{Z} \mathbf Z_{i} + \frac{\sigma^{2}_{X|Z}}{\sigma^{2}_{X|Z} + \sigma^{2}_{U} / N_{i}} (\overline{W}_{i} - (\gamma_{0} + \gamma^{T}_{Z} \mathbf Z_{i}))  \nonumber \\
\mbox{Var}(X_{i}|\mathbf W_{i},\mathbf Z_{i}) &=&  \sigma^{2}_{X|Z} \left(1 - \frac{\sigma^{2}_{X|Z}}{\sigma^{2}_{X|Z}+\sigma^{2}_{U}/N_{i}} \right)
\label{xgivenwz}
\end{eqnarray}
where $\overline{W}_{i}$ denotes the mean of individual $i$'s $N_{i}$ error-prone measurements.

As noted in the introduction, RC gives consistent parameter estimates in the case of a linear outcome model. For logistic and Cox outcome models, RC is approximately consistent. Armstrong first gave justification for RC in generalized linear models under the assumption that $\mbox{Var}(X_{i}|\mathbf W_{i})$ is `small', using the delta method \cite{Armstrong1985}. This condition will be satisfied when the measurement error variance is `small'. Rosner \etal later justified its use in logistic regression under the assumptions that the outcome is rare and that $X_{i}|\mathbf W_{i}$ is normal \cite{Rosner1990}. Subsequently, Kuha showed that RC could be justified as an approximate method for logistic regression provided that $\beta^{2}_{X} Var(X_{i}|\mathbf W_{i})$ is small, without the rare outcome assumption \cite{Kuha1994}. This condition holds when $\beta_{X}$ is small or the measurement error variance is small. For a Cox proportional hazards outcome model, RC can be justified when the event rate is low or the measurement error variance is small \cite{Prentice1982,Hughes1993}.

For valid inferences, the estimation of the parameters involved in $E(X_{i}|\mathbf W_{i},\mathbf Z_{i})$ should be allowed for. One approach is to use bootstrapping. Alternatively, it is possible to construct sandwich variance estimators by stacking the estimating equations used in the two stages \cite{Carroll2006}. One drawback with this approach is that the resulting Wald type symmetric confidence intervals do not reflect the asymmetric sampling distribution of the RC estimator, which may lead to confidence interval coverage which deviates from the nominal level.

\section{Bayesian approach}
\label{bayes}
In this section we describe the key elements of a Bayesian analysis of the covariate measurement error problem.

\subsection{Model specification}
First, we specify a joint parametric model for $(Y_{i},X_{i},W_{i1},W_{i2}|\mathbf Z_{i})$. We condition on the fully observed $\mathbf Z_{i}$, thereby avoiding the need to model its distribution. Assuming that the measurement error is non-differential with respect to both $Y_{i}$ and $\mathbf Z_{i}$, this joint model can be decomposed as
\begin{eqnarray}
f(Y_{i}|X_{i},\mathbf Z_{i},\beta,\eta) f(\mathbf W_{i}|X_{i},\sigma^{2}_{U}) f(X_{i}|\mathbf Z_{i},\gamma) \label{jointmodel}
\end{eqnarray}
The first component is the outcome model, which contains regression parameters of primary interest $\beta=(\beta_{0},\beta_{X},\beta_{Z})$ in the case of linear or logistic regression and $\beta=(\beta_{X},\beta_{Z})$ in the case of Cox regression, and possibly additional parameters $\eta$ (e.g. a residual variance in the case of a linear regression outcome model). The second component is the measurement model, and as described previously the simplest assumption is that the error prone measurements $W_{ij}$ follow a classical error model, with independent normally distributed errors $U_{ij} \sim N(0,\sigma^{2}_{U})$. The final component specifies a model for the unobserved covariate $X_{i}$, conditional on $\mathbf Z_{i}$, with a default choice being a normal linear regression model. We return later to questions of robustness and to model extensions to relax such distributional assumptions. In the case of a Cox proportional hazards outcome model the outcome $Y_{i}$ has two components $(T_{i},D_{i})$ and the additional parameters, $\eta$, denote the baseline hazard function $H_{0}(t)$.

\subsection{Prior specification}
\label{prior.spec}
In the Bayesian approach we must specify priors for the model parameters. The first `Bayesian' analyses made use of flat or constant priors, based on the notion that these represent \textit{a priori} ignorance regarding the value(s) of the model parameter(s) \cite{Berger2006}. The key issue with such priors is that while a flat prior expresses ignorance on one scale, a transformation of the parameter implies a non-flat prior on the transformed parameter. The latter half of the 20th century witnessed the growth of the subjective Bayesian approach, in which the analyst carefully choose the priors to represent their beliefs about the model parameters in advance of seeing the data. Arguably the majority of Bayesian analyses which are now performed by researchers make use of so called non-informative or reference priors \cite{Robert2007}. Such priors do not (and cannot) represent total ignorance about the model parameters, but can be viewed as default priors that one might use when subjective prior information is either not available, or one does not want to use such information in the analysis. The intention of such priors is usually that they have minimal impact on inferences.

For the joint model in equation \eqref{jointmodel}, prior independence is typically assumed for the parameters in the three sub-models. For the outcome model regression coefficients $\beta$ and the coefficients in $\gamma$ a common default prior is a very diffuse normal prior centred at zero. For the variance parameters, the conjugate inverse Gamma distribution has traditionally been advocated. In the context of adjustment for covariate measurement error, Gustafson has proposed using a $Ga(0.5,0.5)$ prior for the precision (reciprocal of variance) parameters \cite{Gustafson2003}. This prior equates to the likelihood that would be obtained from one observation, with the best guess for the precision of one.

Bayesian analysis of the Cox model requires specification of a prior for the baseline cumulative hazard process $H_{0}(t)$ in addition to priors for the regression coefficients $\beta$ and the other submodel parameters. The prior distribution for the baseline cumulative hazard process $H_{0}(t)$ is assumed to be independent of the other priors, including that for $\beta$. Here we use a Gamma process prior for $H_{0}(t)$ as described by Kalbfleisch \cite{Kalbfleisch1978} and Sinha et al. \cite{Sinha2003}, denoted $H_{0}(t)\sim \mathcal{GP}(cH^{*}_{0},c)$, where $H^{*}_{0}(t)$ is a prior guess at the mean and $c$ is a parameter which represents the confidence in that guess, with small values of $c$ corresponding to a diffuse prior. We let $t_{(1)}<t_{(2)}<\cdots <t_{(n^{*})}$ denote the ordered observed event times. Under the assumption that the hazard is degenerate at 0 except at the observed event times $T_{i}$ where $D_{i}=1$ it follows from the Gamma process prior for $H_{0}(t)$ that the increments in the cumulative baseline hazard from time $t_{(j)}$ to time $t_{(j+1)}$ ($j=1,\ldots,n^{*}-1$) have independent Gamma distributions ; $dH_{0}(t_{(j)})\sim \mathrm{Gamma} (c(H^{*}(t_{(j+1)})-H^{*}(t_{(j)}),c)$. In the later application of this approach we use $H^{*}(t_{(j+1)})-H^{*}(t_{(j)}=r(t_{(j+1)}-t_{(j)})$ where $r$ is a guess at the event rate per unit time. It can be shown \cite{Kalbfleisch1978,Sinha2003} that under the Gamma process prior for the cumulative hazard the likelihood for $(\beta,H_{0}(t),c)$ tends to the partial likelihood in the limit as $c$ tends to 0, and that it tends to the full likelihood with $H_{0}(t)=H^{*}$ as $c$ tends to infinity.

\subsection{Posterior inference and simulation}
Given specification of the model and priors, Bayesian inference is then based on the posterior distributions of the model parameters. For the purposes of point estimation the posterior mean is commonly used. To form a 95\% credible interval for a particular parameter, we take the 2.5\% and 97.5\% centiles of the posterior distribution. An advantage of this in the present context of adjustment for covariate measurement error is that asymmetry in the posterior distribution, which typically occurs when adjusting for covariate measurement error, is automatically accounted for in credible intervals.

Except for very specific choices of model and prior, in general the posteriors are not available analytically. Instead, we can utilise Markov Chain Monte Carlo (MCMC) methods to simulate draws from the posteriors distributions (see e.g. part III of \cite{Gelman2004}). The most common approach is the method of Gibbs sampling, in which taking each parameter in turn, a new value is drawn from its full conditional distribution given all other quantities. Often these conditional distributions do not belong to standard parametric families, necessitating the use of more sophisticated sampling techniques (see \cite{Gelman2004,Robert2007} for further details). However, these are implemented in the software packages we describe in the following, such that the analyst need not generally concern themselves with the details.

\subsection{Frequentist properties}
Under certain regularity conditions, as the sample size tends to infinity, the choice of prior has no impact on the posterior distribution, since the latter is then dominated by the likelihood function. Consequently Bayes estimators and uncertainty intervals enjoy the same large sample properties as maximum likliehood methods: the Bayes posterior mean estimator is consistent, asymptotically normal, and efficient \cite{Gelman2004}. In reality of course all samples are finite, and the choice of prior can sometimes have a material affect on inferences. Importantly however, in small samples or sparse data situations, Bayesian methods can have better frequentist properties than ML procedures, particularly if sensible priors are adopted \cite{Greenland2015}.

\subsection{Software}
The explosion of Bayesian data analyses being performed over the last few decades is largely thanks to both the MCMC methods developed and their implementation in acccessible software. Chief among these is the WinBUGS software package, developed in the 1990s \cite{Lunn2009}. It allows the user to define, using a simple language syntax or graphical interface, the model and priors. MCMC methods are then automatically chosen by the package, depending on the specified model and priors. One can then run the MCMC sampler, and after a sufficient number of burnin iterations, draws can be saved as draws from the respective posterior distributions. More recently, new packages have been developed, with developments in various directions. These include the OpenBUGS project (www.openbugs.net) and Stan (mc-stan.org). In the simulations described later, we make use of the Just Another Gibbs Sampler (JAGS) program \cite{Plummer03jags:a}, whose model language is very similar to the BUGS language used by WinBUGS, and which can be easily called from R.

\section{Maximum likelihood and multiple imputation}
\label{mlmi}
\subsection{Maximum likelihood}
Maximum likelihood estimation and inference is based on the likelihood function, but unlike the Bayesian approach, does not involve specification of prior distributions for parameters. Maximum likelihood methods enjoy many favourable frequentist properties - assuming correct model specification the ML estimator is consistent, asymptotically normal, and efficient. In the specific context of adjustment for covariate measurement error, a drawback of ML is that the likelihood function typically involves intractable integrals \cite{Carroll2006}, such that numerical methods such as numerical intergration are required in order to obtain estimates. The same obstacle is overcome in the Bayesian appproach through the use of MCMC methods. A further drawback of ML in the present context is that in small samples inference based on symmetric Wald based confidence intervals may perform poorly due to the lack of regularity of the likelihood function. Software to fit user defined models which allow for covariate measurement error is also somewhat limited. Lastly, the absence of prior distributions prevents the incorporation of external information which may sometimes be available regarding the measurement process.

\subsection{Multiple imputation}
Multiple imputation (MI) has become an extremely popular approach for handling missing data, and has also been advocated as an approach for handling measurement error, in which $X_{i}$ is multiply imputed \cite{Cole2006,Freedman2008,Messer2008,Guo2013}. There is a very close connection between MI and `direct' Bayesian inference. In its originally devised form, MI is based on repeatedly drawing imputations of missing values from their posterior distribution based on a Bayesian imputation model. The analysis model of interest is then fitted, typically using ML, to each of the complete datasets. The resulting estimates and standard errors are then pooled using rules developed by Rubin \cite{Rubin:1987}. MI can most directly be viewed as an approximation to a full Bayesian analysis \cite{Carpenter2013a}, although its frequentist properties can of course also be evaluated \cite{Wang1998}. From the Bayesian perspective, application of MI and Rubin's rules can be viewed as a particular route to performing a Bayesian analysis, in which one effectively assumes that the posterior distributions for the parameters are normally distributed.

As described by Carpenter and Kenward in the context of missing data, there are a number of settings where use of MI may be preferable to a direct Bayesian analysis \cite{Carpenter2013a}. However, in the context of covariate measurement error in parametrically specified outcome models, we argue that the advantages of a direct Bayesian approach far outweigh the disadvantages, relative to MI. First, when only replicate error-prone measurements are available, standard software for performing MI cannot be applied, since $X_{i}$ is missing for all individuals. Second, standard parametric imputation models which might be used to impute $X_{i}$ in general may not be compatible with the assumed outcome model \cite{Bartlett2014}. This will in particular occur when the outcome model is itself non-linear, or the imprecisely measured covariate is assumed to have a non-linear effect on the outcome. Third, when allowance is made for covariate measurement error, as noted earlier, the posterior distributions for the outcome model parameters are typically skewed in small to moderate sample sizes, such that symmetric credible/confidence intervals constructed using Rubin's rules may perform poorly, either from a subjective Bayesian perspective or in a frequentist evaluation. Lastly, we note that software for performing Bayesian inference will typically also permit saving of the imputed values of $X_{i}$ as a by-product, such that if the analyst really wants imputed datasets they can still be obtained.

\section{Simulations}
\label{sims}
In this section we present simulation results for the cases of a linear, logistic and Cox proportional hazards outcome model, comparing the popular RC approach with the Bayesian approach. We adopt the standard frequentist type simulation setup, in which datasets are repeatedly generated using fixed population parameter values.

\subsection{Linear regression}
We first present simulation results for a simple linear regression outcome model. Datasets of size $n=1,000$ were simulated, with covariates $X_{i}$ and $Z_{i}$ drawn from a bivariate normal distribution with means zero, variances one, and covariance 0.25. Continuous outcomes $Y_{i}$ were generated from the linear regression model given in equation \eqref{eq-linear-outcome}, with $\beta_{0}=0$ and $\beta_{X}=\beta_{Z}=1$. The normally distributed residual variance $\sigma^{2}$ was chosen in order to given $R^{2}=0.1,0.5,0.9$. Each individual had an error-prone measurement $W_{i1}=X_{i}+U_{i1}$, with $U_{i1} \sim N(0, \sigma^{2}_{U})$. A random subset of 10\% of individuals had a second error-prone measurement, with the same error variance $\sigma^{2}_{U}$. This variance was chosen to give reliability (unconditional) values of $0.5,0.7,0.9$, corresponding to low, moderate and high reliability.

We first estimated the outcome model parameters using RC, by fitting a random intercepts model for the error-prone measurements, with $Z_{i}$ entering as a fixed effect, as described in Section \ref{rc}. Next we fitted a Bayesian model, calling JAGS from R using the rjags package. We adopted non-informative priors for all model parameters, following those proposed by Gustafson \cite{Gustafson2003}. Specifically we assumed independent normal priors for $\beta_{0},\beta_{X},\beta_{Z},\gamma_{0},\gamma_{Z}$, with mean zero and variance 10,000, and inverse gamma $IG(0.5,0.5)$ priors for each of the variance parameters. As discussed by Gustafson, the latter prior can be thought of as being equivalent to a best guess for the variance of one, coming from a single observation. We ran five parallel chains with 1,000 burnin iterations and 5,000 main iterations. If the Rubin-Gelman convergence statistic Rhat was greater than 1.05 for any of $\beta_{0},\beta_{X},\beta_{Z}$ we extended the chains until this was met.

Table \ref{sim-linear} shows the simulation results, with 1,000 simulations per scenario. RC had slight upward bias for $\beta_{X}$ when the reliability of the error prone measurements was 0.5, and was unbiased for the higher reliability values. The Bayes mean estimator was upwardly biased for reliability equal to 0.5 and $R^{2}=0.1$ and $R^{2}=0.5$. Inspection of the estimates showed that the sampling distribution of the Bayes mean estimator had greater skew than the RC estimator, with the larger estimated values inducing the upward bias. For reliability of 0.5 and 0.7, and $R^{2}=0.9$ the Bayes estimators had lower empirical SD than the RC estimator, and consequently had lower mean squared error in these scenarios. For the other scenarios the RC and Bayes estimators performed similarly. Lastly, the 95\% Bayesian credible intervals had frequentist coverage close to 95\%.

We emphasize that the performance of the Bayesian estimators here depends on the choice of priors. In particular, the use of more informative priors for the measurement error variance and the variance for $X|Z$ could be used to reduce the bias variability of the Bayesian estimators.

\begin{table}[ht]
\caption{Linear regression results. 1,000 simulations per scenario. Empirical means and SDs for estimates of $\beta_{X}$, and coverage of 95\% Bayesian credible intervals.}
\centering
\begin{tabular}{lllll}
  \hline
Reliability & $R^{2}$ & RC & Bayes mean &  Bayes CI \\
  \hline
0.5 & 0.1 & 1.04 (0.29) & 1.17 (0.37)  & 0.94 \\
  0.5 & 0.5 & 1.03 (0.19) & 1.16 (0.28)  & 0.92 \\
  0.5 & 0.9 & 1.02 (0.16) & 0.98 (0.07) & 0.98 \\
  0.7 & 0.1 & 1.01 (0.20) & 1.04 (0.21)  & 0.95 \\
  0.7 & 0.5 & 1.00 (0.09) & 1.03 (0.10)  & 0.95 \\
  0.7 & 0.9 & 1.00 (0.07) & 1.00 (0.05)  & 0.97 \\
  0.9 & 0.1 & 1.01 (0.16) & 1.03 (0.16)  & 0.95 \\
  0.9 & 0.5 & 1.00 (0.06) & 1.02 (0.06)  & 0.95 \\
  0.9 & 0.9 & 1.00 (0.03) & 1.01 (0.03)  & 0.93 \\
   \hline
\end{tabular}
\label{sim-linear}
\end{table}

\subsection{Logistic regression}
Next we performed simulations with a logistic regression outcome model. The covariates $X_{i}$ and $Z_{i}$, and error-prone measurements were generated as described previously. The binary outcome $Y_{i}$ was then generated according to a logistic regression model \eqref{eq-logistic-outcome}. The intercept $\beta_{0}$ was chosen so that $P(Y_{i}=1)=0.2$ approximately. We performed simulations with log odds ratios $\beta_{X}=0.1,0.5,2$, representing small, moderate and large effects of $X_{i}$. As before, we set $\beta_{Z}=\beta_{X}$.

Regression calibration was implemented as described previously. For the Bayesian approach, we again used independent normal priors for each of $\beta_{0}$, $\beta_{X}$ and $\beta_{Z}$. For $\beta_{0}$ we used, as before, the non-informative prior $\beta_{0} \sim N(0,10000)$. For $\beta_{X}$ and $\beta_{Z}$ we adopted the $N(0,1.38)$ prior suggested by Hamra \etal \cite{Hamra2013}. As described by Hamra \etal, the use of such a mildly informative prior can help in terms of stabilizing estimates. This prior corresponds to assuming, \textit{a priori}, that we are 95\% sure that the odds ratios $\exp(\beta_{X})$ and $\exp(\beta_{Z})$ lie between 0.1 and 10, an assumption that is arguably generally reasonable in most epidemiology studies (provided the predictor has been suitably standardized). For the $\gamma$ parameters and the variance parameters we assumed the same priors as in the case of linear regression.

Table \ref{sim-logistic} shows the results of the simulations. For $\beta_{X}=0.1$, RC and Bayes performed very similarly, both being essentially unbiased and having similar empirical SD. For $\beta_{X}=0.5$ and reliability of 0.5, while RC is unbiased, Bayes showed some upward bias and was more variable than RC. For reliability ratios of 0.7 and 0.9 the performance was similar for both. For $\beta_{X}=2$ and reliability of 0.5, RC showed downward bias, consistent with the known properties of RC for logistic regression, in that bias is larger for large covariate effects. In contrast, Bayes showed only a slight downward bias. The bias of RC reduced, again as expected, as the reliability was increased to 0.7 and then 0.9, although some downward bias remained even for the latter case. In contrast Bayes estimates were essentially unbiased. Lastly, the Bayesian 95\% credible intervals had approximately 95\% coverage across all scenarios.

\begin{table}[ht]
\caption{Logistic regression outcome model simulation results, with 1,000 simulations per scenario. Monte-Carlo means and SDs for estimates of $\beta_{X}$ from regression calibration (RC) and Bayes, and empirical coverage of 95\% Bayesian credible intervals}
\centering
\begin{tabular}{lllll}
  \hline
Reliability & $\beta_{X}$ & RC & Bayes mean  & Bayes CI \\
  \hline
0.5 & 0.1 & 0.10 (0.12) & 0.12 (0.14)  & 0.94 \\
  0.5 & 0.5 & 0.51 (0.15) & 0.58 (0.19)  & 0.94 \\
  0.5 & 2 & 1.64 (0.31) & 1.94 (0.32)  & 0.97 \\
  0.7 & 0.1 & 0.10 (0.10) & 0.11 (0.10)  & 0.95 \\
  0.7 & 0.5 & 0.50 (0.11) & 0.52 (0.12)  & 0.94 \\
  0.7 & 2 & 1.74 (0.20) & 2.01 (0.27)  & 0.97 \\
  0.9 & 0.1 & 0.10 (0.09) & 0.10 (0.09)  & 0.94 \\
  0.9 & 0.5 & 0.50 (0.10) & 0.51 (0.10)  & 0.95 \\
  0.9 & 2 & 1.91 (0.17) & 2.01 (0.20)  & 0.96 \\
   \hline
\end{tabular}
\label{sim-logistic}
\end{table}

\subsection{Cox proportional hazards regression}
Lastly, we performed simulations for time-to-event data based on a Cox proportional hazard model. The covariates $X_{i}$ and $Z_{i}$ were generated as described as described for linear regression. Event times $T_{i}$ were generated according to the Weibull hazard model $h(t|X_{i},Z_{i})=\kappa \lambda t^{\kappa-1}e^{\beta_{X} X_{i}+\beta_{Z} Z_{i}}$.  We used $\kappa=2$ and $\lambda$ was chosen so that approximately 10\% of individuals had an event time before the end of follow-up which was fixed at time 10 (e.g. 10 years); the remaining individuals were censored at time 10 ($D_{i}=0$). We performed simulations with log hazard ratios $\beta_{X}=0.1,0.5,2$ and as before we set $\beta_{Z}=\beta_{X}$.

RC was performed as described previously. For the Bayesian approach we used independent normal priors for $\beta_{X}$ and $\beta_{Z}$ and we chose the $N(0,1.38)$ prior as was used in the logistic regression simulations, corresponding here to an assumption that we are 95\% sure that the hazard ratios $\exp(\beta_{X})$ and $\exp(\beta_{Z})$ lie between 0.1 and 10. For the $\gamma$ parameters and the variance parameters we assumed the same priors as in the case of linear and logistic regression. As outlined in Section \ref{prior.spec}, we assumed a process prior for the baseline cumulative hazard which implies Gamma priors for the increments in the hazards; $dH_{0}(t_{(j)})\sim \mathrm{Gamma} (c(H^{*}(t_{(j+1)})-H^{*}(t_{(j)})),c)$. We used $c=0.001$, representing low confidence in the prior mean of the Gamma Process, $H^{*}_{0}(t)$. We used $H^{*}(t_{(j+1)})-H^{*}(t_{(j)})=r(t_{(j+1)}-t_{(j)})$ with $r=0.01$, since the data were simulated so that 10\% of individuals have the event during 10 time units of follow-up. The analysis requires the data to be specified using a counting process format and this is illustrated in example code given online. Due to the higher computational burden of fitting the Cox model, only three parallel chains were used, and 100 (rather than 1000) simulations were performed for each scenario.

\begin{table}[ht]
\caption{Cox regression outcome model simulation results, with 100 simulations per scenario. Monte-Carlo means and SDs for estimates of $\beta_{X}$ from regression calibration (RC) and Bayes, and empirical coverage of 95\% Bayesian credible intervals}
\centering
\begin{tabular}{lllll}
  \hline
Reliability & $\beta_{X}$ & RC & Bayes mean  & Bayes CI \\
  \hline
0.5 & 0.1 & 0.10 (0.09) & 0.10 (0.09)  & 0.98 \\
  0.5 & 0.5 & 0.49 (0.11) & 0.48 (0.11)  & 0.94 \\
  0.5 & 2 & 1.49 (0.15) & 1.92 (0.20)  & 0.92 \\
  0.7 & 0.1 & 0.11 (0.09) & 0.11 (0.09)  & 0.98 \\
  0.7 & 0.5 & 0.49 (0.11) & 0.48 (0.11)  & 0.93 \\
  0.7 & 2 & 1.67 (0.16) & 1.98 (0.18)  & 0.97 \\
  0.9 & 0.1 & 0.11 (0.10) & 0.11 (0.10)  & 0.96 \\
  0.9 & 0.5 & 0.51 (0.10) & 0.50 (0.10)  & 0.96 \\
  0.9 & 2 & 1.84 (0.15) & 1.96 (0.15)  & 0.95 \\
   \hline
\end{tabular}
\label{sim-cox}
\end{table}

Table \ref{sim-cox} shows the simulation results. For $\beta_{X}=0.1$ and $\beta_{X}=0.5$, RC and Bayes performed very similarly across all three reliability values. For $\beta_{X}=2$ and reliability of 0.5, RC showed bias toward the null, in line with previous simulation evidence \cite{Xie2001}. This bias was reduced as the reliability increased, although there was still downward bias with reliability of 0.9. In contrast, the Bayes estimator was much less biased. The Bayesian credible intervals had approximately 95\% coverage across all nine scenarios.

\section{Illustrative example}
\label{example}
To illustrate the potential flexibility and advantages of the Bayesian approach, we consider data from the Third National Health and Nutrition Examination Survey (NHANES III). NHANES III was a survey conducted in the US between 1988 and 1994 in 33,994 individuals aged two months and older. Here we consider an illustrative analyses of the data available on those aged 60 years and above at the time of the original survey. Fitting Cox models to large datasets is very slow using JAGS, particularly for large datasets.  We therefore considered inference for a Weibull regression model for hazard for death due to cardiovascular disease (CVD), with age, sex, smoking status, diabetes status and systolic blood pressure (SBP) at the time of the survey as covariates:
\begin{eqnarray*}
h(t) = r t^{r-1} \exp(\beta_{0} + \beta_{1} \mbox{sbp}_{i} + \beta_{2} \mbox{sex}_{i} + \beta_{3} \mbox{age}_{i} + \beta_{4} \mbox{smoker}_{i} + \beta_{5} \mbox{diabetes}_{i})
\end{eqnarray*}
where $r$ is a shape parameter and $\beta_{1},..,\beta_{5}$ are log hazard ratios. We take the first SBP measurement taken at the survey, $\mbox{sbp}_{i1}$, to be an error-prone measurement of each individual's underlying SBP, subject to classical error. A 5\% subset of individuals was selected to participate in a second examination, during which SBP was again measured. We assume this second measure, $\mbox{sbp}_{i2}$, is an independent error-prone measurement of each individual's underlying SBP. After deleting 7 individuals who were missing diabetes status, data were available on 6,519 individuals. Of these, 1,469 (22.5\%) subsequently died due to CVD, with a median follow-up of 10.8 years. 5,033 (77.2\%) had a first SBP measurement available from the first examination. An SBP measurement at the second examination was available in 401 (6.2\%) of individuals. Unfortunately, smoking status was only recorded in 3,433 (52.3\%) of individuals. The analysis thus required handling of both the measurement error in the SPB measurements and the substantial missingness in the smoking and SBP variables.

\subsection{Naive analyses}
We first fitted the Weibull regression using $\mbox{sbp}_{i1}$ (ignoring measurement error) to the 2,667 complete cases, whose estimates are shown in the first column of Table \ref{nhanesresults}. Strong evidence was found for independent associations between each of the covariates and hazard of death due to CVD, with associations in the expected directions. To check that a Weibull assumption was appropriate, we additionally fitted a Cox proportional hazards model with the same covariates. The estimates of the log hazard ratios between the two models were very similar, suggesting a Weibull assumption is reasonable here. Secondly, we performed a global test of the proportional hazards assumption using the Schoenfeld residuals following fitting the Cox model, which gave $p=0.08$, indicating no evidence to reject an assumption of proportional hazards. Through fitting a logistic regression model for the missingness indicator of the smoking variable, we found evidence that smoking was more likely to be missing for females, older individuals, diabetics, and those individuals with longer follow-up times. The latter finding suggests that the complete case analysis (CCA) may be biased.

Next we fitted the same Weibull model to the complete cases, again ignoring measurement error, using the Bayesian approach. We assumed an exponential prior for the shape parameter $r$ with parameter 0.001. Rather than placing a prior on the log hazard ratios, we placed independent $N(0,10^6)$ priors on $-\beta_{k}/r$, since this leads to improved MCMC mixing \cite{Plummer2015}. Five independent chains were used, with 5,000 burnin iterations and 5,000 main sample iterations. The estimates are 95\% credible intervals are shown in Table \ref{nhanesresults}. In line with theory, due to the large sample size, the Bayesian estimates and intervals were almost identical to those from maximum likelihood.

\subsection{Regression calibration}
We then applied RC to the complete cases. To do this we fitted a linear mixed model to the available SBP measurements, with a random effect for individual and fixed effects of sex, age, smoking and diabetes. We also included a fixed effect to allow for a systematic shift in mean between the first and second exams. The resulting predicted true SBP values at exam one were then used as a covariate to fit the Weibull regression model. We used 2,000 non-parametric bootstrap samples to obtain percentile 95\% confidence intervals for the estimates, in order to take into account the two stage estimation process. Based on the mixed model fit to the complete cases, the estimated reliability (conditional on the error-free covariates) was 0.75. Adjusting for measurement error using RC led to the estimated log hazard ratio for SBP increasing, from 0.085 to 0.115, as expected by approximately $4/3$ (1 divided by the reliability 0.75). Estimates for the other covariates did not materially change.

In order to apply RC to the full dataset, we contemplated use of its use in combination with MI. This is problematic however. First, one could use MI to impute the missing smoking, $\mbox{sbp}_{i1}$ and $\mbox{sbp}_{i2}$ values. For example, one could apply the full conditional specification MI approach, imputing the smoking variable using logistic regression and the SBP variables using linear regression models. In these models one must include the error-free covariates, plus the outcome. In the case of a time to event outcome modelled using a proportional hazards model, an approximately compatible imputation model for covariates includes the event indicator and an estimate of the cumulative hazard function \cite{White2009}. Having generated the imputed datasets, RC could then be applied to each imputed datasets. However, in order to apply Rubin's rules, one requires valid within imputation estimates. As described in Section \ref{rc}, for RC these can only be obtained by bootstrapping or by programming large sample theory estimating equation variance estimators. While both could in principle be programmed, they are not entirely straightforward to implement, and so we do not pursue MI in combination with RC.

\subsection{Bayesian analyses adjusting for covariate measurement error}
Next we modified the Bayesian complete case analysis to accommodate measurement error. We assumed that each individual's true underlying SBP around the time at which the first measurement was obtained, $\mbox{sbp}_{i}$, was normally distributed conditional on smoking, sex, age and diabetes, with $N(0,10^4)$ priors on the regression coefficients and a $Ga(0.5,0.5)$ prior on the precision parameter. For the first SBP measurement, $\mbox{sbp}_{i1}$, we assumed
\begin{eqnarray*}
\mbox{sbp}_{i1} = \mbox{sbp}_{i} + U_{i1}
\end{eqnarray*}
with $U_{i1} \sim N(0,\sigma^{2}_{U})$. For the second SBP measurement, $\mbox{sbp}_{i2}$, we assumed
\begin{eqnarray*}
\mbox{sbp}_{i2} = \nu + \mbox{sbp}_{i} + U_{i2}
\end{eqnarray*}
where $\nu$ is a parameter to allow for a systematic shift in mean between the two exams, and $U_{i2} \sim N(0,\sigma^{2}_{U})$. The errors $U_{i1}$ and $U_{i2}$ are assumed to be independent. For $\sigma^{-2}_{U}$ a $Ga(0.5,0.5)$ prior was assumed, and for $\nu$ a $N(0,10^4)$ prior was assumed. The posterior mean and credible intervals are shown in Table \ref{nhanesresults}, under `Bayes adj. CCA'. The results were very similar to those based on RC CCA, except that the credible interval for SBP was slightly narrower.

A strength of the Bayesian approach is its flexibility to simultaneously handle missing data and measurement error. To accommodate missingness in the smoking and SBP variables under a missing at random assumption, we assumed a model for the distribution of smoking, conditional on the fully observed error free covariates sex, age, and diabetes. In our analysis we assumed a logistic model for this conditional distribution:
\begin{eqnarray*}
\mbox{logit}\left\{P(\mbox{smoker}_{i}=1)\right\} = \alpha_{0} + \alpha_{1} \mbox{sex}_{i} + \alpha_{2} \mbox{age}_{i} + \alpha_{3} \mbox{diabetes}_{i},
\end{eqnarray*}
with independent mean zero normal priors for the regression coefficients, each with variance 10,000. A major advantage of the Bayesian approach here is that the missing smoking and underlying SBP values are imputed by the Gibbs sampler, using the conditional distributions implied by a single well specified joint model for the data. The posterior means were somewhat different to the RC and Bayes complete case analyses, and as expected the credible intervals were narrower, due to the inclusion of observed data from 3,852 individuals. The changes in coefficient estimates may be indicative of bias in the complete case analyses.

\begin{sidewaystable}
\begin{tabular}{llllll}
\hline
Covariate & Naive CCA 											& Naive Bayes						& RC CCA								& Bayes adj. CCA				& Bayes adj. full \\
\hline
SBP (per 20 mmHg) 	& 0.085 (0.014, 0.157) 	& 0.086 (0.015, 0.160)	& 0.115 (0.014, 0.221)	& 0.114 (0.017, 0.211)	& 0.122 (0.059, 0.186) \\
Male 								& 0.49 (0.30, 0.67) 		& 0.49 (0.32, 0.67)			& 0.49 (0.32, 0.68)			& 0.49 (0.31, 0.69)			& 0.46 (0.36, 0.57) \\
Age (per 10 years) 	& 0.88 (0.77, 0.99) 		& 0.87 (0.76, 0.99)			& 0.87 (0.76, 0.99)			& 0.87 (0.75, 0.98)			& 1.01 (0.94, 1.09) \\
Smoker 							& 0.26 (0.07, 0.46) 		& 0.25 (0.06, 0.45)			& 0.26 (0.07, 0.45)			& 0.26 (0.06, 0.46)			& 0.24 (0.07, 0.41) \\
Diabetes						& 0.50 (0.29, 0.72) 		& 0.50 (0.28, 0.72)			& 0.50 (0.28, 0.72)			& 0.50 (0.27, 0.71)			& 0.68 (0.55, 0.81) \\
\hline
\end{tabular}
\caption{Log hazard ratios estimates and 95\% confidence/credible intervals for the NHANES III data. CCA - complete case analysis performed using 2,667 individuals, full analysis performed using 6,519 individuals, RC - regression calibration, naive - ignoring measurement error, adj. - adjusting for measurement error in SBP, SBP - systolic blood pressure.}
\label{nhanesresults}
\end{sidewaystable}

\section{Discussion}
\label{discussion}
In this paper we have empirically compared the frequentist properties of regression calibration and Bayesian approaches to handling covariate measurement error. Our simulations demonstrate that for what might be considered a fairly typical epidemiological study setup, the methods often perform very similarly. As such, we believe that the Bayesian approach for measurement error adjustment may be as useful for the frequentist as for the Bayesian statistician. When the reliability of error-prone measurements was low, the Bayes estimator performed somewhat worse than RC. However, for larger effect sizes, RC was biased for logistic and Cox regression, while the Bayes estimator showed much less bias. A critical point to bear in mind is that there are infinitely many Bayes estimators, corresponding to the different choices of prior distributions - use of different priors could lead to, depending on the true data generating mechanism, better or worse performance. While some analysts dislike the Bayesian approach because of the requirement to specify priors, they give the analyst the opportunity to exploit external information about model parameters, potentially leading to more precise estimates.

We have highlighted the fact that Bayesian estimators enjoy the same large sample frequentist properties as the method of ML, and also described the relationships between these approaches and the popular MI approach. Software for MI cannot be directly applied to handle covariate measurement error when replication data are available. Moreover, even when validation data are available, the covariate imputation models included in MI implementations may not be compatible with the analyst's outcome model \cite{Bartlett2014}. A further strength of the Bayesian approach is that uncertainty intervals automatically allow for the skewness typically found in covariate measurement error adjusted estimators.

As has been noted by many authors before, a key strength of the Bayesian approach is its flexibility to handle more complicated models and data structures. As we have demonstrated in Section \ref{example}, the Bayesian approach can readily accommodate both covariate measurement error and missing data. Moreover, more complex measurement error models can in principle be used, for example to allow for heteroscedastic error, systematically biased measurements, or more flexible modelling of the true covariate's distribution \cite{Carroll2006}.  The flexibility of the Bayesian approach also lends itself to the problem of adjusting for covariate measurement error when the true covariate is assumed to have a complex non-linear association with the outcome \cite{Berry2002}. In this paper we have focused on the setting whereby internal replication data are available; the Bayesian approach readily handles the situation where validation data are instead available.

Nonetheless, the Bayesian approach has a number of drawbacks. As a fully parametric approach, a natural concern is sensitivity of inferences to distributional assumptions, particularly those about the unobserved true covariate and measurement errors. In this regard the Bayesian approach can utilize more flexible model specifications, for example by modelling the unobserved true covariate using a normal mixture model \cite{Richardson2002}. An important practical issue is that although the software available for fitting complex analyst defined models using the Bayesian approach has seen dramatic developments over the last 25 years \cite{Lunn2009}, fitting certain models (e.g. Cox proportional hazards models) can still take tremendously long. Although this concern will be progressively mitigated by increasing computational power, it is arguably still a material drawback. Further research and effort is therefore warranted to develop software implementations of the Bayesian approach which mitigate this.

\section*{Supplementary material}
R and JAGS code demonstrating each of the simulation setups, and code and data for the illustrative analysis, are provided at the GitHub repository \url{https://github.com/jwb133/bayesMeasurementError}.

\section*{Funding}
This paper was written while J.W.B. was a member of the Department of Medical Statistics, London School of Hygiene and Tropical Medicine, and was supported by a Medical Research Council Fellowship  [MR/K02180X/1]. R.H.K. was supported by a Medical Research Council Fellowship [MR/M014827/1].

\end{document}